# Multimodal EEG and Keystroke Dynamics Based Biometric System Using Machine Learning Algorithms

ARAFAT RAHMAN[1], MUHAMMAD E. H. CHOWDHURY[2*], AMITH KHANDAKAR[2,3], SERKAN KIRANYAZ[2], KH SHAHRIYA ZAMAN[3], MAMUN BIN IBNE REAZ[3], MOHAMMAD TARIQUL ISLAM[3], MAYMOUNA EZEDDIN[2], MUHAMMAD ABDUL KADIR[1]

[1]Department of Biomedical Physics & Technology, University of Dhaka, Dhaka-1000, Bangladesh
[2]Department of Electrical Engineering, Qatar University, Doha 2713, Qatar
[3]Department of Electrical, Electronic and Systems Engineering, Universiti Kebangsaan Malaysia, Bangi 43600, Selangor, Malaysia

* **Correspondence:** Muhammad E. H. Chowdhury, mchowdhury@qu.edu.qa

**ABSTRACT** Electroencephalography (EEG) based biometric systems are gaining attention for their anti-spoofing capability but lack accuracy due to signal variability at different psychological and physiological conditions. On the other hand, keystroke dynamics-based systems achieve very high accuracy but have low anti-spoofing capability. To address these issues, a novel multimodal biometric system combining EEG and keystroke dynamics is proposed in this paper. A dataset was created by acquiring both keystroke dynamics and EEG signals simultaneously from 10 users. Each user participated in 500 trials at 10 different sessions (days) to replicate real-life signal variability. A machine learning classification pipeline is developed using multi-domain feature extraction (time, frequency, time-frequency), feature selection (Gini impurity), classifier design, and score level fusion. Different classifiers were trained, validated, and tested for two different classification experiments – personalized and generalized. For identification and authentication, 99.9% and 99.6% accuracies are achieved, respectively for the Random Forest classifier in 5 fold cross-validation. These results outperform the individual modalities with a significant margin (~5%). We also developed a binary template matching-based algorithm, which gives 93.64% accuracy 6X faster. The proposed method can be considered secure and reliable for any kind of biometric identification and authentication.

## I. INTRODUCTION

Numerous advancements in technology have enabled real-time applications through remote access in banking, healthcare, military, retail, enterprise, law enforcement, and many other sectors. However, in such sensitive applications, security is a vital aspect, which usually comes at the cost of convenience. Over the years, many types of authentication tokens have been established, such as a shared secret token, look-up token, out-of-bound token, event-based one-time password (OTP) token, cryptographic token, biometric token, and hybrid/multi-modal token [1]. Biometric systems play an important role in the identification and authentication infrastructure. Biometric systems utilize the unique physiological and behavioral characteristics of an individual to generate authentication tokens. Nevertheless, there are several vulnerabilities in these methods due to the use of the same password for multiple accounts, brute force attacks, key logger attacks, etc. Even biometric systems are susceptible to spoofing attacks through fabrication or falsification. This is done by using photos for face recognition, recordings for voice recognition, lift tape for fingerprints, and high-resolution pictures for iris scanners, etc.

The biometric authentication systems that rely on Electroencephalogram (EEG) signals from the brain extract unique behavioral patterns to identify the user. EEG signals are obtained by recording electrical wave patterns at different parts of the brain, which is done by capturing the EEG signals from the brain using invasive or non-invasive techniques. The invasive technique, also considered as an intracranial EEG or Electrocorticography (ECoG) requires surgery to implant special needle electrodes inside the brain while the non-invasive technique simply requires placing small electrodes on the scalp [2]. Such systems have many benefits: (i) they are robust against forgery and spoofing attacks as the signal acquisition is currently impossible through a remote interface and the signal depends on the state of the mind; (ii) the EEG signal has been shown to possess low variance within an individual and high variance among a group of individuals; (iii) it exists in all living people; (iv) different working states can be gathered which can be valuable for restricting authentication in the non-natural state; (v) EEG can be collected spontaneously from users without certain specific actions. However, authentication systems that solely rely on EEG signals achieve low accuracy and are prone to instability over time using multi-channel signals [3]. On top of that, EEG has a low spatial resolution on the scalp, and a weak signal-to-noise ratio (SNR) [4]. The emotional states at different sessions play a crucial role in EEG-based biometric systems. Arnau-González et al. [5] reported that identification accuracy can vary greatly due to different emotions.

Keystroke dynamics is a reliable and easy-to-implement biometric system that does not require additional hardware to acquire data. Keystroke dynamics analyses a user's typing patterns by recording strokes and gaps between keyboard inputs. The data are then matched with a learned pattern from the manner and rhythm of the authentic user. Keystroke dynamics produce distinctive features for each individual and can produce a very high identification rate. However, it is prone to spoofing attacks through falsification as it can be easily seen during typing and the typing pattern can be easily mimicked later.

In this paper, we propose a novel multi-modal authentication and identification system that combines EEG and keystroke dynamics with robust machine-learning algorithms. Unlike the previously proposed multimodal authentication systems, our model integrates EEG signals and keystroke dynamics into the most widely used PIN/password infrastructure. To the best of our knowledge, this is the first attempt to combine EEG with keystroke dynamics. The EEG data and keystroke dynamics were collected simultaneously while a user was typing a specific password. The EEG signals were acquired using a portable 5 channel Emotiv Insight EEG system, which is user-friendly, easy-to-setup, and a comparatively cheap EEG solution. The dataset was created with different age and gender groups to fairly evaluate the proposed multimodal system. Such a dataset is also a contribution of this research, as there are no publicly available datasets with these two modalities. Several machine learning techniques were implemented on the fused multimodal dataset, and the top 7 best-performing algorithms were evaluated. Furthermore, we also evaluated generalized and personalized models on the individual modality and the fused modalities. While the generalized models were a multi-class problem, the personalized models were trained as a two-class classifier for each user. Finally, a real-time framework with a graphical user interface (GUI) including a template matching approach was also developed.

The paper has three major contributions. Firstly, we collected a dataset combining EEG and keystroke at 10 different sessions from 10 subjects of different ages and genders. To the best of our knowledge, this is the first attempt to build this type of dataset. Secondly, we established a robust algorithmic pipeline that achieves very high accuracy for the combination of EEG and keystroke and this algorithm overcomes the limitation of individual modalities. Thirdly, we proposed a method to build efficient binary templates for fast and secured authentication in real-time. We tested our dataset and algorithm in different scenarios like closed set generalized and personalized classification for all the modalities (EEG, keystroke, and the combination of EEG and keystroke) with different feature sets and augmentation techniques. We also reported the Cumulative Matching Characteristic (CMC) curve for template matching in identification scenarios and the Equal Error Rate (EER) with Receiver Operating Characteristic (ROC) curve for authentication scenarios where our method achieved better performance for the fusion of EEG and keystroke than individual modality.

The rest of the paper is organized as follows. In Section II, the related works of this research domain are presented with comparative advantages and disadvantages. In Section III, the methodology for the proposed multimodal system, which includes experimental setup, data preparation, and processing are described. The experimental results are presented in Section IV and discussed in Section V. Finally, we leave some concluding remarks in Section VI.

## II. RELATED WORKS

Several works have been done using only EEG for biometric authentication. Bai et al. [6] developed a system using the EEG signals related to visual evoked potential combined with different feature reduction techniques like Genetic Algorithm, Fisher Discriminant Ratio, and Recursive Feature Elimination and achieved an accuracy of 97.25% using Support Vector Machine (SVM). Phung et al. [7] used Shannon's Entropy-based features and achieved an accuracy of 94.9%. Gui et al. [8] used Euclidean and Dynamic Time Warping (DTW) based matching technique and achieved an accuracy of 80% and 68% respectively. Zhang et al. [9] used single-channel EEG to identify 46 subjects using Rayleigh quotient feature selection and ensemble classifier with an accuracy of 95.48%. Yang et al. [10] proposed a method that utilizes the logarithm of wavelet coefficients and discrete cosine transform. Ruiz-Blondet et al. [11] utilized a discriminant function-based classifier on the event-related potential (ERP) of 50 users to achieve an accuracy of 100% in the identification scenario. Nakamura et al. [12] used a novel user-friendly EEG biometric system that can be placed inside the ear and used combined power spectral density (PSD) and autoregressive coefficients (AR) with linear discriminant analysis (LDA) to achieve an accuracy of 87.2%. Fraschini et al. [13] used functional connectivity of the brain to identify persons. Moctezuma et al. [14] designed an ERP-based identification system where they proposed a four objective optimization method and achieved 97.02% accuracy. Sabeti et al. [15] compared different feature sets and showed that correlation and spectral coherence-based features produce the best results with an accuracy of 97.36% and 97.08% respectively, for resting-state EEG in the verification scenario. However, most of the works discussed above used EEG systems that require a large number of

VOLUME XX, 2021    9

channels like 32, 64, or 128, which is bulky, requires an hour or more to set up, and computationally expensive.

Several studies used keystroke dynamics to improve the verification and authentication process. Wang et al. [16] proposed a method named differential evolution and adversarial noise-based user authentication (DEANUA) for reducing equal error rate (EER) in smartphone-based keystroke biometrics. They extracted 146 features and selected the best features from that set using differential evolution. They also tried to increase the robustness of their system by synthesizing adversarial noise samples. Saevanee and Bhatarakosol [17] proposed a user authentication system that uses key-hold time, inter-key duration, and finger pressure as features and achieved 1% EER using finger pressure. Giot et al. [18] proposed a keystroke dynamics-based user authentication method that uses the hold time of a key, inter-key time as features.

Several researchers have explored the multi-modal authentication system by combining EEG and other biometrics to add an extra layer of security. Saini et al. [19] and Kumar et al. [20] combined users' signatures with the EEG signals generated during the signature for identifying the user. They used machine learning techniques to extract features from the handwritten signature and the simultaneously recorded EEG signals. Using their multi-modal framework, they have achieved higher authentication accuracy compared to that of individual biometric systems. Wang et al. [21] proposed an authentication system where they combined event-related EEG signals with a facial recognition algorithm. They have extracted features from EEG and facial data independently and then used cross-correlation to fuse the modality and produce a combined identification system. The authors suggested that by implementing deep neural networks, they could improve the accuracy of the multimodal system. Zhang et al. [22] developed a multimodal authentication system based on a recurrent neural network (RNN), called DeepKey, where they used EEG and gait signals in parallel to identify the user. They achieved a low false rejection rate (FRR) and false acceptance rate (FAR) of 0% and 1% respectively. Krishna et al. [23] incorporated EEG signal acquisition along with eye movement tracking in an augmented reality (AR) and virtual reality (VR) headset for user authentication. Bashar [24] explored the combination of brain signals (EEG) and heart signals (ECG) for authentication using wavelet domain statistical features with multiple classifiers and achieved a 90.5% F1 score. Ibtehaz et al. [25] proposed a framework called EDITH, which performs competitively using just a single ECG heartbeat (96- 99.75% accuracy) and can be further enhanced by fusing multiple beats (100% accuracy from 3 to 6 beats).

## III. METHODOLOGY

The proposed multimodal authentication system incorporated the user's keystroke dynamics and brain signals as illustrated in Fig. 1. The system setup and data acquisition are described below:

### A. SYSTEM SETUP AND DATA ACQUISITION

This study was conducted using a personal computer for keystroke dynamics acquisition and Emotiv INSIGHT headset with 5 sensors + 2 references, a Bluetooth-based wireless headset [26] for EEG signal acquisition. A specific password was asked to be typed in the keyboard for each user repetitively 50 times to capture keystroke dynamics, and EEG (brain) signals are captured using the Emotiv INSIGHT device non-invasively. The letter in the user input was used as the start and end markers for the pre-processing and classification of EEG signal can be carried out on each password-typing event.

Then the captured EEG signal is sampled at 128 Hz using Xavier Test Bench software from Emotiv that displays real-time EEG signal, sensors' contact impedance, and battery level. The users were instructed to wear the headset on their head while keeping their eyes open and active and not to be involved in any other cognitive tasks. In this headset, EEG electrodes are oriented based on an international 10-20 system. The entire recording session is monitored where the contact impedance is greater than 1 kΩ, but less than 10 kΩ [27] that ensures similar signal throughput from all channels. EEG signals are recorded from temporal, parietal, frontal, and occipital (AF3, AF4, T7, T8, PZ) lobes of the user to ensure that the system is capturing the different important lobes of the human brain. Keystroke dynamics and EEG data were collected from 10 users for 10 sessions (days) and 50 trials per session. The user was typing a specific password

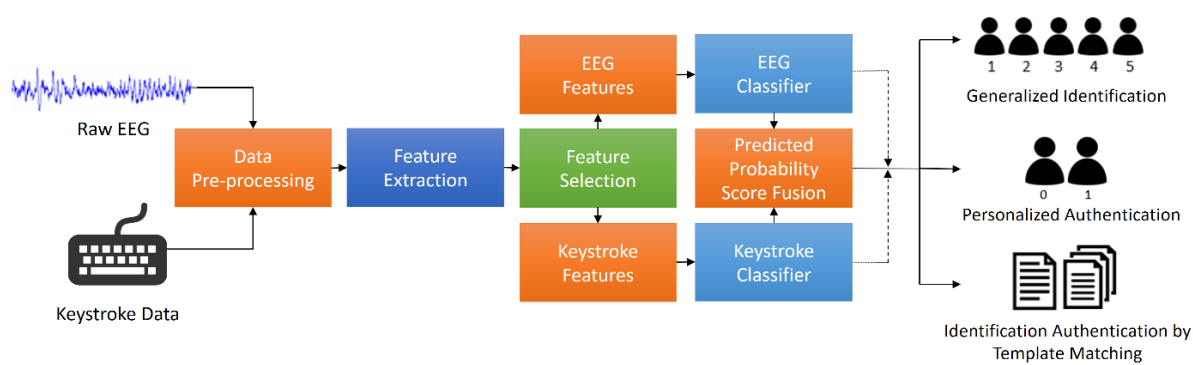

**FIGURE 1.** Flow diagram for the proposed data acquisition and multimodal authentication method for EEG and Keystroke dynamics.

"qu-ELEC371" 50 times while the EEG and keystroke dynamics were recorded. Therefore, the EEG dataset consists of 5000 samples, where each user contributes 500 samples. Since the data were acquired on 10 different days and therefore, involuntary motion artifacts, random noises, and other subjects' psychological and physiological variability were present in the different sessions. It was noticed that in the different sessions, severe motion artifacts and noises were present in the dataset. In summary, it can be said that it is a well-representation of real-life data.

For the keystroke dynamics, the raw data consisted of a time-trail of keystrokes for the characters of the password. An in-house written code was used to captures different time points of keystrokes in-synchronization to the EEG signal acquisition and extracted the required features (durations) from the timestamps. The study was approved by the ethical board of Qatar University (QU-IRB) and written consent was obtained from participants to join in the research. There were five male and five female subjects (mean age: 25 years, standard deviation: 20 years) who participated in this study.

### B. DATA PREPROCESSING

EEG signal is inherently very weak in amplitude and despite different efforts to keep the acquisition environment quiet and the EEG signal motion artifacts minimum, some noises and signal deviation were observed in the recorded EEG signal. Therefore, the raw EEG signal was undergone several pre-processing steps to mitigate such errors. The signal processing steps are outlined below:

*Baseline correction:* The baseline drift introduces unwanted amplitude shifts in the signal and this leads to poor performance in the classification model. A 6-degree polynomial curve fitting algorithm was used to approximate the baseline and it is subtracted from the raw signal to correct the baseline.

*Filtering:* Bandpass filtering of the EEG signal is very important to ensure the band-limited EEG signals are used for further analysis. The bandwidth of the Emotiv Insight EEG system is 0.5-63 Hz and therefore, a band-pass infinite impulse response (IIR) filter with a cutoff frequency of 0.5 Hz and 63 Hz was used for removing any potential unwanted noises.

*Segmentation:* As mentioned earlier, the starting and ending time-stamp of a password entry were marked in the EEG data. In-house-built MATLAB code was used to automatically identify the start and end markers and extract the EEG signal segment corresponding to a password entry.

*Resampling:* Different users take a different amount of time to type a full password and also typing duration can vary from trial to trial. Therefore, segmented EEG signal length can vary for different trials. However, the feature extraction process and machine learning models require equal sample EEG length for any trial independent of session or subjects. Therefore, all the signal samples are resampled to 1024 samples. This signal length is determined by observing the mean signal length in the entire dataset.

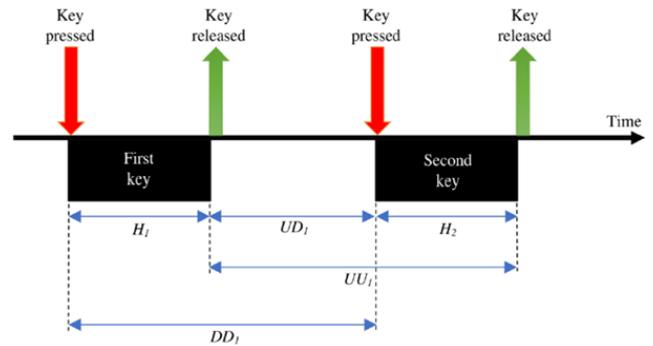

**FIGURE 2.** Features of keystroke dynamics, representing key-hold time (H), key-up to key-up time (UU), key-down to key-down time (DD), and key-up to key-down time (UD).

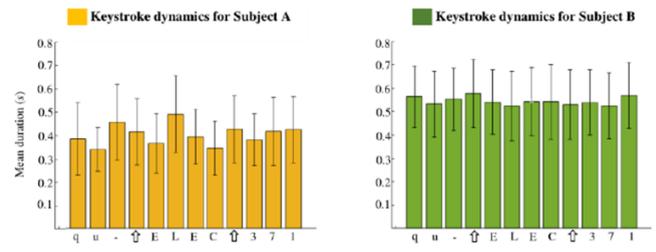

**FIGURE 3.** Comparison of key-hold time (H) for two random subjects.

### C. FEATURE EXTRACTION

Since the EEG signals and keystroke data were collected for 10 subjects while typing the password 50 times in 10 different sessions, it was possible to mimic the inter-session variability for the same subject while providing enough training data for machine learning models. For efficient training of classifiers and to extract meaningful information from raw data, feature extraction is necessary. Multimodal features (from keystroke dynamics and EEG signals) were extracted and used to train different machine learning models to identify the user in generalized and personalized experiments.

#### 1) KEYSTROKE DYNAMICS

Four different features were calculated from the keystroke timestamps: key-hold time (H), key-down to key-down time (DD), key-up to key-up time (UU), and key-up to key-down time (UD) for each keypress, as shown in Fig. 2. The unique password used for all users was "qu-ELEC371" to show that even though everyone will type the same password, the machine learning model can still authenticate the correct user. This resulted in a total of 500 samples per user over multiple sessions. Each sample contains data for 12 key-presses including two caps lock (⇧) (one for activating capital letter mode and another for deactivating). Therefore, for each user, there are 45 features (11 features for UU, DD, and UD each, and 12 features for H). Fig. 3 illustrates and compares the key-hold time (H) features for two random subjects. As seen from Fig. 3, the H features may be stable for one subject (Subject B), while the H time for each keypress may vary for another user (Subject A). It was also noted that the special character '-' had large standard deviations for most users. This is most probably because it is a less frequently used character on the keyboard. The typing speed and typing rhythm of each user resulted in a unique set of features for each subject.

2) EEG

For EEG signal, 13-time domain, 10 frequency domain, and 18 time-frequency features based on the feature reported in the literature were calculated for each channel and a feature called Signal Magnitude Area Total (SMAT) was calculated combining five channels. Therefore, the total number of features from the EEG signal is 206, where five times 41 plus one SMAT feature. The range of values in different channels of raw EEG signals can vary greatly due to several reasons. So, to bring all the features to the same range, all the features were normalized to a value between 0 and 1. All the features with their definitions and relevant formulas are shown below in TABLE I.

TABLE I
EEG FEATURES, DEFINITION, AND MATHEMATICAL EXPRESSION

| Features | Definition | Mathematical Expression |
|---|---|---|
| **Mean** | Mean is the sum of all data divided by the number of entries. | $\bar{x} = \frac{\sum_{i=1}^{n} x_i}{n}$ (1) <br><br> $x_i$ is the $i^{th}$ data sample and $n$ is the total number of samples. |
| **Median** | The median is in the middle data of the ordered set of data. | Median = middle number if n is odd <br><br> = mean of two middle numbers if n is even |
| **Standard Deviation** | Standard deviation is the measure of variability and consistency of the sample. | $s = \sqrt{\frac{\sum_{i=1}^{n}(x_i - \bar{x})^2}{n-1}}$ (2) |
| **Mean Absolute Deviation (MAD)** | Mean Absolute Deviation (MAD) is the average distance between the mean and each data value. | $MAD = \frac{\sum_{i=1}^{n}|x_i - \bar{x}|}{n}$ (3) |
| **25th Percentile** | 25th Percentile is the data value at which the percent of the value in the data set is less than or equal to 25. | $25^{th} = \left(\frac{25}{100}\right)n$ (4) |
| **75th Percentile** | 75th Percentile is the data value at which the percent of the value in the data set is less than or equal to 75. | $75^{th} = \left(\frac{75}{100}\right)n$ (5) |
| **Inter Quartile Range (IQR)** | Inter Quartile Range (IQR) is the measure of the middle 50% of a data set. | $IQR = Q3 - Q1$ (6) <br><br> Here, $Q1$, $Q3$ are the first and third quartiles, respectively. |
| **Skewness** | Skewness is the measure of the lack of symmetry from the mean of the dataset. | $g = \frac{\sum_{i=1}^{n}(x_i - \bar{x})^3/n}{s^3}$ (7) |
| **Kurtosis** | Kurtosis is the pointedness of a peak in the distribution curve, in other words, it is the measure of the sharpness of the peak of the distribution curve. | $k = \frac{\sum_{i=1}^{n}(x_i - \bar{x})^4/n}{s^4} - 3$ (8) |
| **Hjorth Activity** | Hjorth Activity represents the variance of the amplitude of the signal. It is part of Hjorth Parameters introduced by [28]. It can indicate signal power. | $Hact = var(x(t))$ (9) <br><br> Here, $x(t)$ is the the time-varying amplitude of the signal. |
| **Hjorth Mobility** | Hjorth Mobility represents the mobility of the signal, which can indicate mean frequency. From the power spectrum, the | $Hmob = \sqrt{\frac{var(x'(t))}{var(x(t))}}$ (10) |
| | proportion of standard deviation of power can be also inferred from this feature. It is also part of Hjorth Parameters. | Here, $x'(t)$ is the 1st derivative of the amplitude of the signal. |
| *Hjorth Complexity* | Hjorth Complexity represents the complexity, which can specify the change of the signal frequency. It is also part of the Hjorth Parameters. | $Hcom = \frac{Hmob(x'(t))}{Hmob(x(t))}$ (11) |
| *Shannon's Entropy* | Shannon's Entropy measures the degree of randomness in a set of data and higher entropy indicates greater randomness, and lower entropy indicates lower randomness. | $H(X) = -\sum_{i=0}^{n} p(x_i) \log_2 p(x_i)$ (12) Here, $X$ is a discrete random variable, $x_i$ is the $i^{th}$ outcome of $X$, $p(x_i)$ = Probability of occurring $x_i$ |
| *Spectral Entropy (SEN)* | Spectral Entropy (SEN) is the normalized Shannon entropy that is applied to the power spectrum density of the signal. | $SEN = \frac{-\sum_{i=o}^{N-1} p_k \log_2 p_k}{\log N}$ (13) Here, $p_k$ = the spectral power of the normalized frequency, $N$ = the number of frequencies in binary. |
| *Second Max Frequency (M2F)* | The second Maximum Frequency (M2F) is the frequency at which the 2nd maximum amplitude of the frequency spectrum occurs. | |
| *Second Max Frequency Amplitude* | The second Maximum Frequency Amplitude is the 2nd maximum amplitude of the frequency spectrum. | |
| *Second Max Relative Energy* | Second Max Relative Energy is the summation of energy between *M2F+δ* to *M2F-δ*. Then the summation was normalized by total energy below the cutoff. Here, *δ* has a constant value of 5 Hz. | |
| *Average Band power* | The average band powers of alpha, beta, gamma, theta, delta, and raw EEG signal. For this purpose, at first raw EEG signal is decomposed into alpha, beta, gamma, theta, delta bands, and then 6 band powers of EEG bands including the band power of raw EEG were computed as features. | Delta (0-4Hz), Theta (4-7Hz), Alpha (7-13Hz), Beta (13-30Hz) and gamma (30-63Hz) and raw (0-63Hz) |
| *Wavelet Features* | Discrete wavelet decomposition was used to decompose the raw EEG signal into 18 frequency bands. Then average band powers of all of those bands were calculated. Here, order 8 Daubechies wavelet is used as a mother wavelet. | $\gamma_{jk} = \int_{-\infty}^{\infty} x(t) \frac{1}{\sqrt{2^j}} \psi\left(\frac{t-k2^j}{2^j}\right) dt$ (14) Here, $\psi(t)$ is the mother wavelet that is shifted at $j$ and scaled at $k$ and $x(t)$ is the original signal. |

### D. FEATURE SELECTION

To avoid overfitting and to get rid of irrelevant and superfluous features, feature selection is necessary. Therefore, the correlation coefficients between the features were calculated and highly correlated features ($r_{xy} > 0.95$) were removed from any correlated pair. After removing highly correlated features, the remaining features were ranked according to the importance score obtained from the feature ranking model. Random Forest is an ensemble of decision trees, which was used for the feature ranking. It calculates feature ranking using "Gini impurity" or "mean decrease impurity" [29]. The importance scores of a feature at all the nodes of a decision tree are calculated and then averaged over all trees. In a decision tree with child nodes, while partitioning at node *m*, the decrease in impurity at node *m* can be calculated by the following equation,

$$\Delta i_m = i_m - \sum_{j=1}^{a} w_j i_j \quad (15)$$

Here, $\Delta i_m$ is the decrease of impurity at node *m*, $i_m$ is the impurity value at node *m*, $w_j$ is the weighted number of samples at node *j*, $i_j$ is the impurity value at node *j*, and *a* is the total number of child nodes. Impurity at node *m* is calculated by "Gini Impurity Index" which is defined by the following formula,

$$i_m = \sum_{q=1}^{C} P(q) * (1 - P(q)) \quad (16)$$

Here, $C$ is the total number of classes, $P(q)$ is the probability of selecting a data-point of class $q$. Then the importance of a feature $x_k$ is calculated by averaging the decrease of impurity at all nodes over all trees as shown in the following formula,

$$Imp(x_k) = \frac{1}{T} \sum_{t=1}^{T} \sum_{m=1}^{N} w_m \Delta i_m \quad (17)$$

Here, $\Delta i_m$ is the decrease of impurity at node $m$, $w_m$ is the weighted number of samples at the $m^{th}$ node, $N$ is the total number of nodes where the feature $x_k$ appears, and $T$ is the total number of trees.

After ranking the features according to their importance, the performances of the classifiers were evaluated with respect to the number of features, and the performance saturation and over-fitting were observed to identify the best feature combinations. The same procedure was followed for the keystroke data.

### E. IDENTIFICATION AND AUTHENTICATION
#### 1) IDENTIFICATION BY CLASSIFICATION
In the identification setup, we classified 10 subjects using EEG, Keystroke, and the combination of EEG and Keystroke data. This type of classification is also known as generalized classification. We validated and tested several machine learning classifiers for identifying 10 subjects. A 5 fold random stratified cross-validation experiment was conducted to validate the proposed method. In this experiment, the data were split into 5 folds and the models were trained, validated, and tested 5 times. During each training time, the proportion of train and test set was 80:20. 20% of the training set was used for validation. In this manner, the train, validation, and test sets are changed alternatively 5 times. The reported performance metrics are average accuracy (A), precision (P), sensitivity/recall (R), and F1 score (F1) of the test set.

#### 2) AUTHENTICATION BY CLASSIFICATION
In the authentication setup, we set the problem as a two-class (genuine, imposter) classification problem. For this purpose, we have treated one subject as the genuine class and the other nine as the imposter class. This scheme of the dataset was also categorized into train, validation, and test folds using 5-fold cross-validation. In this manner, we have tested all the subjects and reported the average test scores. This setup is also called personalized classification. While treating one subject as genuine and others as imposters, a class imbalance problem arises which reduces the performance of the classifiers. As there are very few training samples of the minority class, the classifier struggles to learn and detect the minority class. To solve this problem, we applied several data augmentation algorithms on the training data, such as

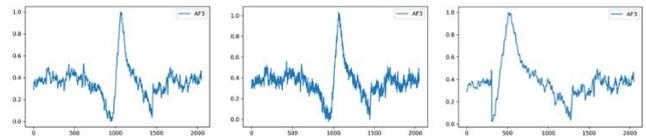

**FIGURE 4.** Comparison of different types of augmentation applied to an EEG segment of channel AF3 - Original raw EEG (left); after Jitter augmentation (middle) and after TimeW augmentation (right).

Jitter, Time Warping (TimeW), Synthetic Minority Oversampling Technique (SMOTE), and Adaptive Synthetic Sampling (ADASYN). Brief descriptions of these algorithms are given below:

***Jitter:*** In Jitter augmentation, we tried to simulate the addition of noise in the synthetic data. For that purpose, a small amount of Gaussian noise was added without altering the actual labels [30]. So, the Jitter augmentation of a signal can be defined by the following equation,

$$X'_n = X_n + J_n \quad (18)$$

Here, $X'_n$ is the augmented $n^{th}$ data point, $X_n$ is the actual $n^{th}$ data point of a signal segment, $J_n \sim \mathcal{N}(0, \sigma)$ is a random noise point drawn from the Gaussian distribution of 0 mean and standard deviation of $\sigma$.

Using this strategy, we synthesized data for the genuine class to make the samples equal to the imposter class. The noise was added equally in 5 channels of the raw EEG data. Here, we used a Gaussian distribution of 0 mean and standard deviation, $\sigma = 0.05$ to generate noise samples. It should be noted that this type of augmentation is only applicable to raw EEG data.

***Time Warping (TimeW):*** TimeW is an augmentation technique that randomly modifies the temporal position of a signal [31]. It simulates the variation of the temporal location of an event in a time window. At first, data was segmented into $N$ temporal slices (where $N$ varies from 2 to 4) and a random slice was chosen. Then the signal in that time range is randomly shifted followed by a random compression or expansion. TimeW can be defined by the following equation,

$$X'(t) = X(Z_n t + Z_n) \quad (19)$$

Here, $X'(t)$ is the augmented signal, $X(t)$ is the actual signal segment, $Z_n \sim \mathcal{N}(1, \sigma)$ is a random noise point drawn from the Gaussian distribution of mean = 1 and standard deviation, $\sigma = 0.05$. Fig. 4 shows the comparison of Jitter and TimeW.

***Synthetic Minority Oversampling Technique (SMOTE):*** Synthetic Minority Oversampling Technique (SMOTE) is an oversampling technique that increases the minority class samples by synthesizing new samples on the line drawn between several random existing samples [32]. In order to generate new samples, at first, this algorithm randomly selects a minority sample and its k nearest neighbors. Then

VOLUME XX, 2021                                                                                                                                                            9

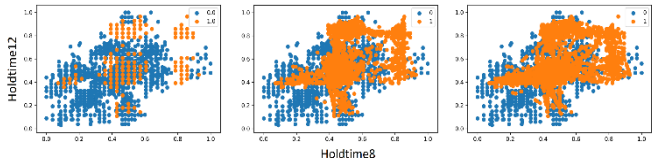

**FIGURE 5.** Scatter plot of Holdtime8 and Holdtime12 features without augmentation (left), after SMOTE augmentation (middle) and after ADASYN augmentation (right) from Keystroke data. Blue dots represent the imposter class whereas orange ones represent the genuine class.

it draws lines between the selected samples on the feature space. Finally, new examples are generated along those lines.

*Adaptive Synthetic Sampling (ADASYN):* To increase the minority samples, we used another technique called Adaptive Synthetic Sampling (ADASYN). This algorithm simultaneously focuses on the region where the density of the minority class is low and harder to learn [33]. Unlike SMOTE, which generates samples randomly, ADASYN generates new samples adaptively in the mentioned regions and this strategy gives better performance. SMOTE and ADASYN both works on the extracted feature and not on the raw signals like Jitter and TimeW. Fig. 5 shows the comparison of different augmentation techniques applied to feature space.

### 3) IDENTIFICATION BY TEMPLATE MATCHING

For real-life identification, we analyzed our system's performance using a template matching method. In this method, at first, we calculated some binary templates from the extracted feature vectors of EEG and Keystroke data. The binary templates are suitable as they enable us to perform very quick Hamming distance calculation for template matching. It also allows various template encryption methods for security [34], [35]. To quantize the real feature values in the range [0, 1], we have calculated a binary template, $B(k)$ of a feature $k$ using the following equation,

$$B(k) = \begin{cases} 1, & \text{if } x_k \geq t_k \\ 0, & \text{otherwise} \end{cases} \quad (20)$$

Here, $x_k$ is the real value of feature $k$, $t_k = median(R_k)$, where $t_k$ is the threshold for feature $k$ and $R_k$ is the list of all values of feature $k$.

After quantization, the train sets were considered as the gallery and test sets as the probe. For each subject in the gallery, there were several samples and all the samples were grouped. Thus, there were 10 groups for 10 subjects. Then for each sample in the probe, Hamming distances were calculated between the probe and all the samples of a gallery group. Then the lowest Hamming distance was selected from all the distances and later placed in a matching matrix. This procedure was repeated for 10 groups. Finally, a matching matrix of $P \times G$ was constructed, where $P$ is the number of samples in the probe and $G$ is the number of groups in the gallery. This matrix can be illustrated by the following equation,

$$M = \begin{bmatrix} m_{11} & m_{12} & \cdots & m_{1g} \\ m_{21} & m_{22} & \cdots & m_{2g} \\ \vdots & \vdots & \ddots & \vdots \\ m_{p1} & m_{p2} & \cdots & m_{pg} \end{bmatrix} \quad (21)$$

After constructing the matching matrix, all the elements in a row were sorted in an ascending manner according to their values and it was repeated for all rows. Then the matching scores were replaced with their corresponding subject index. Finally, the matrix was evaluated using the Cumulative Match Characteristic (CMC) curve and Rank (N) recognition rate [36]. So, for each query in the probe set, an algorithm ranked all the gallery samples according to their Hamming distances in the ascending order, and the CMC Rank(N) identification rate is measured as follows,

$$CMC(N) = \frac{1}{P} \sum_{i=1}^{P} \begin{cases} 1, & \text{if top } N \text{ gallery samples contain the query sample} \\ 0, & \text{otherwise} \end{cases} \quad (22)$$

Here, $P$ is the total number of samples in the probe set. Using this formula, identification rates for $N = 1, 2, 3, \ldots, 10$ were evaluated and a graph of identification rate vs rank was drawn. Finally, the identification rates for Rank(1), Rank(2), and Rank(3) were reported.

### 4) AUTHENTICATION BY TEMPLATE MATCHING

In the authentication scenario, there are only two classes (genuine and imposter). So, the matching matrix has only 2 columns. In this setup, a sample in the probe set is considered genuine if its genuine matching score is less than the imposter's matching score. In this manner, predictions for probes are generated. Finally, this method is evaluated using False Acceptance Rate (FAR), False Rejection Rate (FRR), and Equal Error Rate (EER). Here, FAR and FRR are defined by the following formulas,

$$FAR = \frac{Number\ of\ rejected\ genuines}{Total\ number\ of\ genuines} \quad (23)$$

$$FRR = \frac{Number\ of\ accepted\ imposters}{Total\ number\ of\ imposters} \quad (24)$$

For different threshold values, FAR and FRR are calculated and two curves of FAR vs threshold and FRR vs threshold are drawn. In that graph, the intersection point of two curves is observed and the error value in that point is considered as EER. So, EER is the value where FAR is equal to FRR. Finally, EER for each subject and average EER are also calculated.

All the experiments were carried out using a PC with Intel(R) Core(TM) i5-4200 CPU with a clock speed of 1.6 GHz, 4 GB RAM, and a Windows 10 professional 64 bit. All the pre-processing and feature extraction was done using MATLAB 2018b and classifiers were trained and tested using different Python libraries like Scikit-learn [37], NumPy [38], pandas [39], Matplotlib [40], etc.

## F. SCORE LEVEL FUSION

Two separate classifiers were trained and tested for EEG and Keystroke data. The classifiers give probability scores for each class during prediction and the scores are summed using the following formula,

$$S = \sum_{i=1}^{2} P_i(m_i = c) \qquad (25)$$

Here, $P_i(m_i = c)$ is the probability of assigning the identity $c$ to the person in modality $m_i$. By using this formula the score for each class is calculated and the highest performing class is taken as the final prediction. This strategy can be also named "Late Fusion". Other fusion strategies like – early fusion (feature fusion), Canonical Correlation Analysis (CCA), Discriminant Correlation Analysis (DCA), and multiplying the probability scores were also tested but they did not give satisfactory performance. So, the sum of the probability scores was taken as the best method.

## IV. EXPERIMENTAL RESULTS

This section demonstrates the performance of different experiments conducted in this study. Firstly, the important feature selection results for EEG and Keystroke data were reported. The remaining section is divided into several subsections where we showed the experimental results achieved by our system in closed set identification and authentication. We showed the results of several classifiers in identification or generalized classification and then subject-wise authentication or personalized classification for all modalities. Then, the identification and authentication results using the proposed template matching method are also shown. Finally, a comparison of inference time for different methods is reported.

### A. FEATURE IMPORTANCE

Fig. 6 shows the feature importance plot for EEG and Keystroke data. The most important feature for EEG is PZ_cD2, which is the band power of the level 2 wavelet decomposition of the PZ channel (parietal channel). It can be noticed that the other important features are also from wavelet decomposition (PZ_cD3, T7_cD2, PZ_cD4, etc.). For Keystroke, the most important feature is downdown8, which is the key-down to key-down time of character 8 or the key-down to key-down time between 'C' and 'Caps Lock' keys. Fig. 7 shows the performance vs the number of features for EEG and Keystroke data. It is observed that the performance starts to saturate at 54 features for EEG and at 34 features for Keystroke data. So, finally, 54 and 34 features are selected for EEG and Keystroke, respectively.

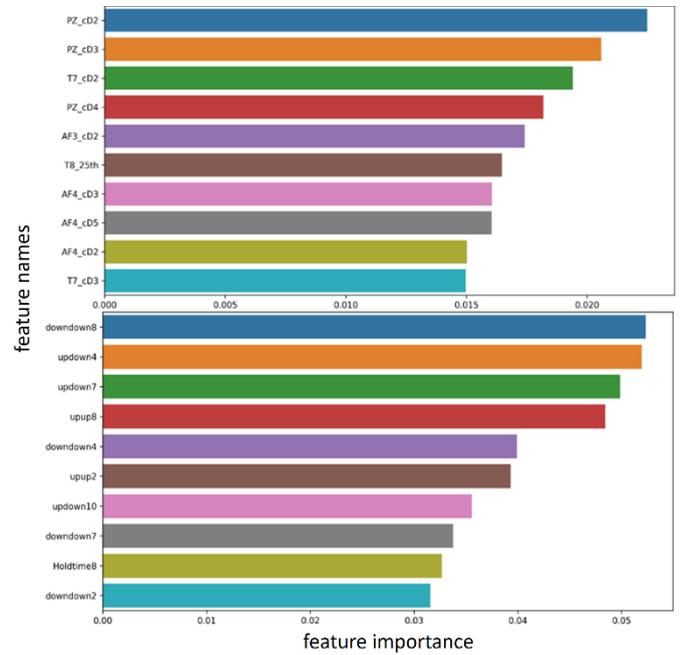

FIGURE 6. Top-10 features using Random Forest feature ranking model for EEG signal (top) and Keystroke data (bottom).

### B. GENERALIZED IDENTIFICATION

Several machine learning algorithms were trained with a single modality and fused-modality for identifying 10 subjects. Stratified 5-fold cross-validation was used for performance evaluation. The test scores of 7 top classifiers for EEG with and without feature selection are shown in TABLE II.

*EEG:* It can be noticed from TABLE II that the highest accuracy, precision, recall, and F1 score are obtained by Random Forest (RnF) classifier. RnF obtained 95.8%

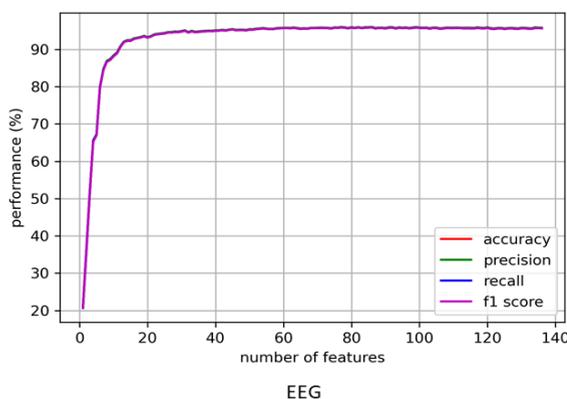

EEG

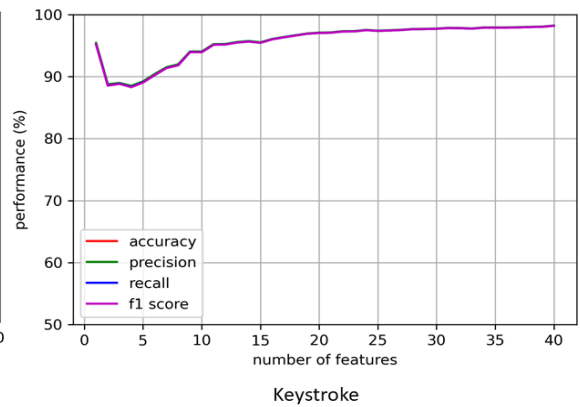

Keystroke

accuracy, 96.0% precision, 95.8 % recall and 95.8% F1 score. The results are

TABLE II
COMPARISON OF TOP 7 CLASSIFIERS USING EEG AND KEYSTROKE FOR IDENTIFYING 10 SUBJECTS

| Strategy | Classifier | EEG | | | | Keystroke | | | | EEG and Keystroke | | | |
|---|---|---|---|---|---|---|---|---|---|---|---|---|---|
| | | A | P | R | F1 | A | P | R | F1 | A | P | R | F1 |
| All features | LSVM | 86.8 | 87.3 | 86.7 | 86.9 | 92.6 | 92.7 | 92.5 | 92.6 | 95.4 | 95.3 | 95.2 | 95.2 |
| | QSVM | 91.6 | 91.7 | 91.5 | 91.6 | 95.4 | 95.3 | 95.2 | 95.2 | 96.5 | 96.5 | 96.5 | 96.5 |
| | KNN | 86.4 | 86.6 | 86.3 | 86.4 | 86.4 | 86.7 | 86.4 | 86.5 | 90.0 | 90.4 | 90.0 | 90.1 |
| | CART | 78.6 | 78.9 | 78.6 | 78.5 | 87.1 | 87.1 | 87.1 | 87.0 | 86.8 | 86.8 | 86.8 | 86.7 |
| | XGBoost | 93.0 | 93.3 | 93.0 | 93.0 | 99.3 | 99.3 | 99.3 | 99.3 | 99.5 | 99.5 | 99.5 | 99.5 |
| | RnF | **95.8** | **96.0** | **95.8** | **95.8** | **99.5** | **99.3** | **99.3** | **99.3** | **99.9** | **99.9** | **99.9** | **99.9** |
| | LDA | 78.8 | 79.8 | 78.7 | 79.2 | 97.9 | 97.9 | 97.9 | 97.9 | 99.5 | 99.3 | 99.3 | 99.3 |
| with feature selection | LSVM | 81.4 | 82.1 | 81.4 | 81.7 | 91.1 | 91.3 | 91.2 | 91.2 | 94.0 | 94.0 | 94.1 | 94.0 |
| | QSVM | 91.4 | 91.4 | 91.2 | 91.2 | 94.8 | 94.9 | 94.9 | 94.9 | 96.0 | 96.0 | 96.0 | 96.0 |
| | KNN | 88.1 | 88.4 | 88.3 | 88.3 | 85.9 | 86.1 | 85.9 | 85.9 | 89.5 | 89.7 | 89.6 | 89.6 |
| | CART | 79.0 | 79.4 | 79.0 | 78.6 | 87.0 | 87.1 | 87.0 | 86.9 | 88.2 | 88.5 | 88.2 | 88.2 |
| | XGBoost | 91.4 | 91.8 | 91.4 | 91.4 | 98.9 | 98.9 | 98.9 | 98.9 | 99.0 | 99.0 | 99.0 | 99.0 |
| | RnF | **95.8** | **96.0** | **95.8** | **95.8** | **99.1** | 98.9 | 98.9 | 98.9 | 99.4 | 99.3 | 99.1 | 99.2 |
| | LDA | 64.1 | 66.2 | 64.0 | 65.0 | 98.2 | 98.2 | 98.2 | 98.2 | **99.6** | **99.6** | **99.6** | **99.6** |

*highest score is indicated by bold numbers

consistent in both the cases - with and without feature selection. The best hyper-parameters of RnF are chosen by random search algorithm (number of trees = 500, split criterion = 'Gini', min samples split = 2, and max features is the square root of the total number of features). The second-best performing classifier is Extreme Gradient Boosting (XGBoost), which achieved 93% accuracy, 93.3% precision, 93% recall, and 93% F1 score without feature selection. XGBoost achieved slightly lower scores with feature selection. However, it is evident that ensemble algorithms with bagging and boosting techniques are the most suitable ones for EEG biometric classification.

*Keystroke:* Same classifiers were tested on the Keystroke data as shown in TABLE II. Here, we can see that RnF again achieved the highest accuracy, precision, recall, and F1 score of 99.5%, 99.3%, 99.3%, 99.3%, respectively without feature selection. RnF also achieved the highest performance with feature selection; however, the performance metrics slightly reduce than that were reported with all features. The second-best performing classifier is XGBoost which achieved 99.3% accuracy, 99.3% precision, 99.3% recall, and 99.3% F1 score. This analysis clearly shows that Keystroke can achieve better performance than EEG. Fusion of EEG and Keystroke can improve the performance metrics even further.

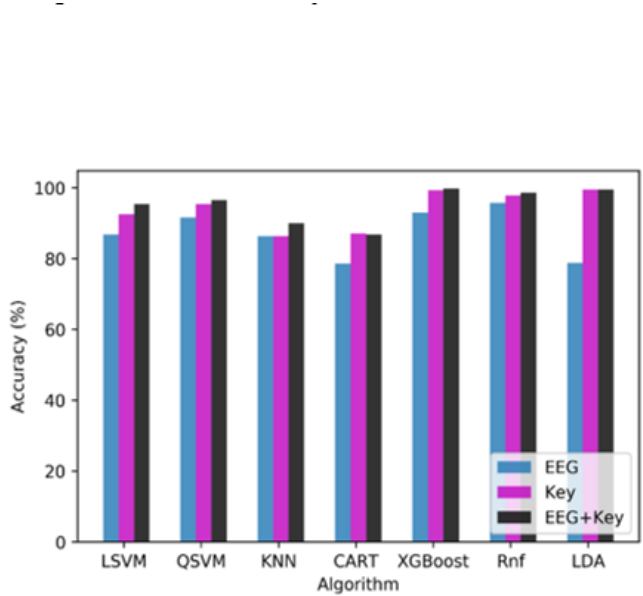

FIGURE 7. Performance vs the number of features pl...

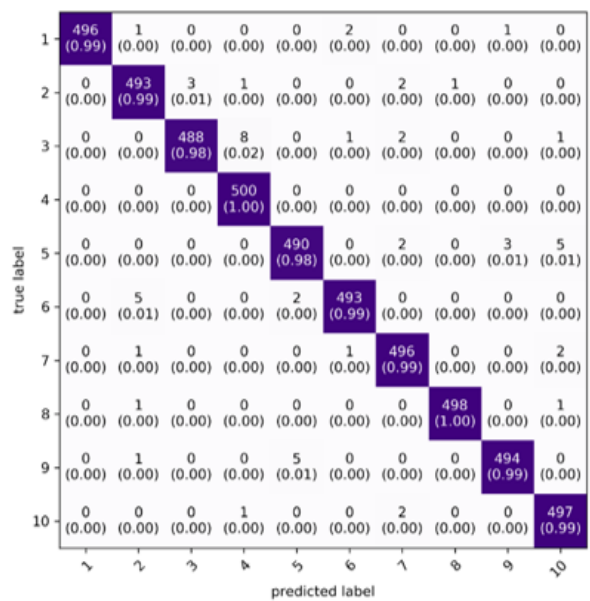

FIGURE 8. Comparison of overall accuracy for the 7 classifiers (left) and co... using the combination of EEG

*EEG and Keystroke:* The achieved results using the fused modality are also shown in TABLE II. It can be seen from TABLE II that the RnF achieved the highest accuracy, precision, recall, and F1 score of 99.9%, 99.9%, 99.9%, and 99.9%, respectively without feature selection. But with feature selection, LDA achieved the highest accuracy, precision, recall, and F1 score of 99.6%, 99.6%, 99.6%, and 99.6%, respectively among all the classifiers. A comparison of fused modality with EEG reveals that fused modality achieved 4.1% higher accuracy, 3.9% higher precision, 4.1% higher recall, and 4.1% higher F1 score than EEG. Fusing the techniques marginally improved performance over Keystroke alone, and the result is improved by 0.4%, 0.6%, 0.6%, and 0.6% for accuracy, precision, recall, and F1 score, respectively. Therefore, it is evident from these experiments that the fused modality always performs better than the individual modalities. Fig. 8 shows the performance comparison of the classifiers using independent modality and fused modalities using a bar chart and the confusion matrix for the fused multimodal dataset. The model was able to identify subject 4 with the highest sensitivity and the most confusing subjects are subjects 3 and 5. Supplementary Fig. 1 shows the macro averaged Receiver operating characteristic (ROC) curve for the best classifier using independent modality and fused modalities.

## C. PERSONALIZED AUTHENTICATION

Several classifiers were evaluated using stratified 5-fold cross-validation for classifying genuine and imposter classes in different modalities. Random Forest performed the best in terms of average accuracy, precision, recall, and F1 score. TABLE III shows the test scores for EEG, Keystroke, and a combination of them for all the subjects using the RnF classifier.

*EEG:* From Table III, it can be noticed that the average accuracy, precision, recall, and F1 score without feature selection are 98.3%, 98.7%, 91.7%, and 94.6%, respectively. Using feature selection, the average scores remain almost identical. Though accuracy and precision remain high for all the classes, recall and F1 scores drop in some classes. This is the consequence of a huge class imbalance between the genuine and imposter classes. In this experiment, the ratio of genuine and imposter class is 1:9. As a result, the classifier fails to learn genuine class examples due to a small number of representative data. To get rid of this problem, we applied two types of data augmentation on the raw EEG signal and they are Jitter and TimeW. Supplementary TABLE I shows the results with Jitter and TimeW augmentations. It can be seen that the average accuracy, precision, recall, and F1 score with TimeW are 98.42%, 98.85%, 92.25%, and 95.12%, respectively. The average accuracy, precision, recall, and F1 score with Jitter are 98.58%, 98.61%, 93.38%, and 95.70%, respectively. Though TimeW has shown almost no improvement, Jitter shows a little improvement of 0.42% in average recall and 0.15% in average F1 score. If we look at the individual subject level performance, we can see Jitter improves the recall scores of subjects 3, 7, 8, 9, and 10 by 3.3%, 2.3%, 1.6%, 1.3%, and 8.6% respectively.

TABLE III
COMPARISON OF PERSONALIZED AUTHENTICATION IN EEG, KEYSTROKES, AND EEG-KEYSTROKES WITHOUT AUGMENTATION

| Strategy | Subject | EEG | | | | Keystroke | | | | EEG and Keystroke | | | |
|---|---|---|---|---|---|---|---|---|---|---|---|---|---|
| | | A | P | R | F1 | A | P | R | F1 | A | P | R | F1 |
| All features | 1 | 99.8 | 99.0 | 99.9 | 99.5 | 99.8 | 99.9 | 99.0 | 99.4 | 99.8 | 99.9 | 99.0 | 99.4 |
| | 2 | 96.4 | 98.1 | 82.1 | 88.0 | 99.0 | 99.5 | 95.1 | 97.1 | 98.6 | 99.2 | 93.0 | 95.9 |
| | 3 | 98.4 | 99.1 | 92.2 | 95.2 | 99.4 | 99.7 | 97.2 | 98.3 | 98.4 | 99.1 | 92.0 | 95.2 |
| | 4 | 99.6 | 99.8 | 98.1 | 98.9 | 98.6 | 99.2 | 93.0 | 95.9 | 99.8 | 99.9 | 99.0 | 99.4 |
| | 5 | 98.4 | 99.1 | 92.2 | 95.2 | 99.4 | 99.7 | 97.1 | 98.3 | 98.8 | 99.3 | 94.0 | 96.5 |
| | 6 | 97.6 | 98.7 | 88.0 | 92.5 | 99.4 | 99.7 | 97.0 | 98.3 | 99.4 | 99.7 | 97.0 | 98.3 |
| | 7 | 98.8 | 99.3 | 94.3 | 96.5 | 99.4 | 99.7 | 97.0 | 98.3 | 99.6 | 99.8 | 98.0 | 98.9 |
| | 8 | 98.4 | 98.1 | 92.9 | 95.3 | 99.0 | 99.5 | 95.0 | 97.1 | 99.4 | 99.7 | 97.0 | 98.3 |
| | 9 | 98.8 | 99.3 | 94.3 | 96.5 | 99.6 | 99.8 | 98.0 | 98.9 | 99.4 | 99.7 | 97.0 | 98.3 |
| | 10 | 96.4 | 96.7 | 82.9 | 88.3 | 99.6 | 99.8 | 98.1 | 98.9 | 99.6 | 99.8 | 98.0 | 98.9 |
| | **Average** | **98.3** | **98.7** | **91.7** | **94.6** | **99.3** | **99.7** | **96.7** | **98.0** | **99.3** | **99.6** | **96.4** | **97.9** |
| with feature selection | 1 | 99.6 | 98.9 | 98.9 | 98.9 | 99.8 | 99.9 | 99.0 | 99.4 | 99.8 | 99.9 | 99.0 | 99.4 |
| | 2 | 97.0 | 98.4 | 85.0 | 90.4 | 99.0 | 99.5 | 95.0 | 97.1 | 98.8 | 99.3 | 94.0 | 96.5 |
| | 3 | 98.4 | 99.1 | 92.0 | 95.2 | 99.4 | 99.7 | 97.0 | 98.3 | 99.2 | 99.6 | 96.0 | 97.7 |
| | 4 | 99.8 | 99.9 | 99.1 | 99.4 | 98.8 | 99.3 | 94.0 | 96.5 | 99.4 | 99.7 | 97.0 | 98.3 |
| | 5 | 98.8 | 99.3 | 94.2 | 96.5 | 99.2 | 99.6 | 96.0 | 97.7 | 99.2 | 99.6 | 96.0 | 97.7 |
| | 6 | 97.6 | 98.7 | 88.1 | 92.5 | 99.6 | 99.8 | 98.0 | 98.9 | 99.6 | 99.8 | 98.0 | 98.9 |
| | 7 | 98.4 | 99.1 | 92.1 | 95.2 | 99.8 | 99.9 | 99.0 | 99.4 | 99.2 | 99.6 | 96.0 | 97.7 |
| | 8 | 98.2 | 97.0 | 92.8 | 94.8 | 99.2 | 99.6 | 96.0 | 97.7 | 99.4 | 99.7 | 97.0 | 98.3 |
| | 9 | 99.0 | 99.5 | 95.0 | 97.1 | 99.6 | 99.8 | 98.0 | 98.9 | 99.2 | 99.6 | 96.0 | 97.7 |

| | 10 | 96.8 | 96.9 | 84.9 | 89.8 | 99.2 | 99.6 | 96.1 | 97.7 | 99.4 | 99.7 | 97.0 | 98.3 |
| | **Average** | **98.4** | **98.7** | **92.2** | **94.9** | **99.4** | **99.7** | **96.8** | **98.2** | **99.3** | **99.7** | **96.6** | **98.1** |

TABLE IV
PERFORMANCE OF PERSONALIZED CLASSIFICATION AFTER AUGMENTATION ON FEATURE VECTORS OF EEG, KEYSTROKES, AND EEG-KEYSTROKES

| Augmentation | Subject | EEG | | | | Keystroke | | | | EEG and Keystroke | | | |
|---|---|---|---|---|---|---|---|---|---|---|---|---|---|
| | | A | P | R | F1 | A | P | R | F1 | A | P | R | F1 |
| SMOTE | 1 | 99.6 | 98.1 | 99.8 | 98.9 | 99.8 | 99.9 | 99.0 | 99.4 | 100.0 | 100.0 | 100.0 | 100.0 |
| | 2 | 98.4 | 95.6 | 95.6 | 95.6 | 98.6 | 99.2 | 93.0 | 95.9 | 99.0 | 99.4 | 95.0 | 97.0 |
| | 3 | 98.8 | 98.4 | 94.9 | 96.5 | 99.0 | 97.6 | 96.8 | 97.2 | 99.6 | 98.8 | 98.8 | 98.8 |
| | 4 | 100.0 | 100.0 | 100.0 | 100.0 | 99.6 | 98.9 | 98.9 | 98.9 | 99.8 | 99.0 | 99.8 | 99.4 |
| | 5 | 99.4 | 99.7 | 97.0 | 98.3 | 99.4 | 97.9 | 98.8 | 98.4 | 99.8 | 99.8 | 99.0 | 99.4 |
| | 6 | 98.8 | 99.3 | 94.0 | 96.5 | 99.8 | 99.0 | 99.9 | 99.5 | 99.8 | 99.0 | 99.8 | 99.4 |
| | 7 | 99.4 | 99.7 | 97.0 | 98.3 | 99.4 | 97.9 | 98.8 | 98.4 | 99.6 | 99.7 | 98.0 | 98.8 |
| | 8 | 98.4 | 95.6 | 95.6 | 95.6 | 100 | 100 | 100 | 100 | 99.6 | 98.8 | 98.8 | 98.8 |
| | 9 | 99.6 | 99.8 | 98.1 | 98.9 | 99.6 | 99.8 | 98.0 | 98.9 | 99.8 | 99.8 | 99.0 | 99.4 |
| | 10 | 97.8 | 94.2 | 93.4 | 93.8 | 99.4 | 98.8 | 97.9 | 98.3 | 99.4 | 97.9 | 98.7 | 98.3 |
| | **Average** | **99.0** | **98.0** | **96.5** | **97.2** | **99.5** | **98.9** | **98.1** | **98.5** | **99.6** | **99.2** | **98.7** | **98.9** |
| ADASYN | 1 | 99.4 | 97.2 | 99.7 | 98.4 | 99.8 | 99.9 | 99.0 | 99.4 | 100.0 | 100.0 | 100.0 | 100.0 |
| | 2 | 98.4 | 96.3 | 94.7 | 95.5 | 98.6 | 99.2 | 93.0 | 95.9 | 99.0 | 98.4 | 95.8 | 97.1 |
| | 3 | 99.0 | 97.6 | 96.8 | 97.2 | 98.8 | 98.4 | 94.9 | 96.5 | 99.6 | 99.7 | 98.0 | 98.8 |
| | 4 | 99.8 | 99.0 | 99.9 | 99.4 | 99.2 | 98.6 | 96.9 | 97.7 | 100.0 | 100.0 | 100.0 | 100.0 |
| | 5 | 99.6 | 99.8 | 98.0 | 98.9 | 99.4 | 97.9 | 98.8 | 98.4 | 99.8 | 99.8 | 99.0 | 99.4 |
| | 6 | 98.6 | 97.3 | 94.8 | 96.0 | 99.6 | 98.9 | 98.9 | 98.9 | 99.6 | 98.8 | 98.8 | 98.8 |
| | 7 | 98.8 | 96.7 | 96.7 | 96.7 | 99.4 | 97.9 | 98.8 | 98.4 | 99.8 | 99.8 | 99.0 | 99.4 |
| | 8 | 98.2 | 94.0 | 96.33 | 95.1 | 100 | 100 | 100 | 100 | 99.6 | 98.8 | 98.8 | 98.8 |
| | 9 | 99.8 | 99.9 | 99.0 | 99.4 | 99.6 | 99.8 | 98.0 | 98.7 | 99.8 | 99.8 | 99.0 | 99.4 |
| | 10 | 98.0 | 94.4 | 94.4 | 94.4 | 99.2 | 98.6 | 96.9 | 97.7 | 99.6 | 98.8 | 98.8 | 98.8 |
| | **Average** | **98.9** | **97.2** | **97.0** | **97.1** | **99.4** | **98.9** | **97.5** | **98.2** | **99.6** | **99.4** | **98.7** | **99.1** |

To improve the sensitivity of EEG-based modality further, we have applied SMOTE and ADASYN augmentation on the extracted features from the EEG data. The results of these augmentations are shown in TABLE IV. It can be seen from Table IV that the average accuracy, precision, recall, and F1 score with SMOTE are 99%, 98%, 96.5%, and 97.2%, respectively. The average accuracy, precision, recall, and F1 score with ADASYN are 98.9%, 97.2%, 97%, and 97.1% respectively. There is a significant improvement in sensitivity that can be observed after using SMOTE and ADASYN. For SMOTE and ADASYN, the improvement is 4.3% and 4.8%, respectively. It is evident that ADASYN performs the best among the augmentation techniques. In the case of individual subject scores, significant improvements can be seen in subjects 2, 4, 6, and 10.

***Keystroke:*** The performance metrics of Keystroke data for personalized classification is also reported in Table III. It can be seen from Table III that the average accuracy, precision, recall, and F1 score without feature selection are 99.3%, 99.7%, 96.7%, and 98.0%, respectively. The average performance metrics did not improve significantly using feature selection.

Hence, to increase the overall performance, different augmentations techniques were applied. SMOTE and ADASYN were applied on the Keystroke feature vectors and the results are reported in Table IV. It can be noticed from TABLE IV that the average accuracy, precision, recall, and F1 score with SMOTE are 99.5%, 98.9%, 98.1%, and 98.5%, respectively. On the other hand, the average accuracy, precision, recall, and F1 score with ADASYN are 99.4%, 98.9%, 97.5%, and 98.2%, respectively. The best average recall score is obtained by applying SMOTE augmentation which is 1.4% higher than without any augmentation.

***EEG and Keystroke:*** Finally, we combined EEG and Keystroke and tested with and without augmentation. Table III shows the results of the fused modality with and without feature selection. It can be seen from Table III that a significant performance improvement in average metrics was achieved after combining EEG and Keystroke. The average accuracy, precision, recall, and F1 score without feature selection are 99.3%, 99.6%, 96.4%, and 97.9%, respectively. Feature selection did not significantly improve the performance. The sensitivity of fused modality is 4.7% and 4.4% higher than EEG alone for without and with feature selection, respectively. Due to the class imbalance, recall value was suffering here too. Thus, we applied different augmentations on the composite feature vectors. Table IV shows the results obtained after augmentation on the composite feature vectors of EEG and Keystroke data. The average accuracy, precision, recall, and F1 score with SMOTE are 99.6%, 99.2%, 98.7%, and 98.9%, respectively. On the other hand, the average accuracy, precision, recall, and F1 score with ADASYN are 99.6%, 99.4%, 98.7%, and 99.1%, respectively. In comparison to other methods, it is evident that ADASYN performs the best.

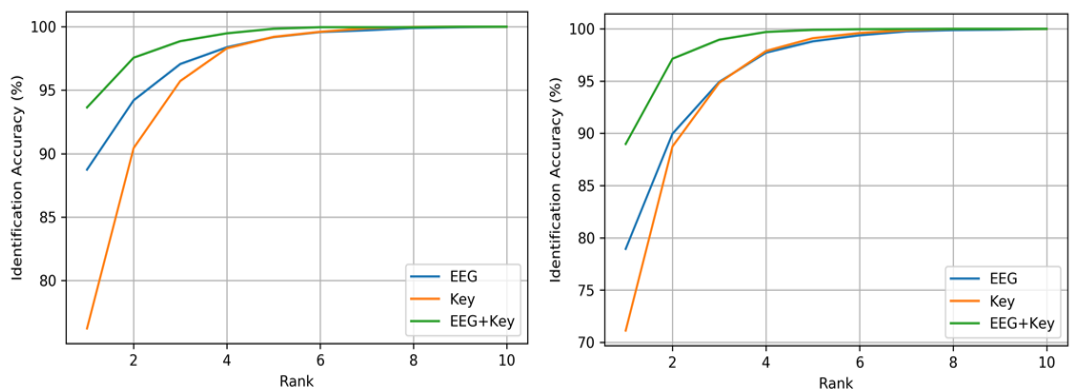

**FIGURE 9.** CMC curves obtained for EEG, Keystroke and EEG-Keystroke modalities: without feature selection (left) and with feature selection (right).

TABLE V
RANK WISE IDENTIFICATION RATES FOR EEG, KEYSTROKES AND COMBINED MODALITIES

| Strategy | Rank | Recognition rate (%) (EEG) | Recognition rate (%) (Keystroke) | Recognition rate (%) (EEG and Keystroke) |
|---|---|---|---|---|
| All features | 1 | 88.74 | 76.22 | **93.64** |
|  | 2 | 94.20 | 90.44 | **97.56** |
|  | 3 | 97.06 | 95.72 | **98.86** |
| with feature selection | 1 | 78.94 | 71.12 | **88.98** |
|  | 2 | 89.96 | 88.74 | **97.14** |
|  | 3 | 94.92 | 94.84 | **98.96** |

### D. IDENTIFICATION BY TEMPLATE MATCHING

As previously stated, a matching matrix is calculated for identifying 10 subjects and from that matrix, CMC curve, Rank(1), Rank(2), and Rank(3) recognition rates are obtained. Fig. 9 shows the CMC curve for EEG, Keystroke, and the combination of EEG and Keystroke. TABLE V shows Rank (1), Rank (2), and Rank (3) recognition rates for all modalities. It is obvious from Fig. 9 that the fused modality outperforms the individual modalities by a significant margin. It is also evident that EEG performs better than Keystroke. It can be seen from Table V that the Rank (1) accuracy of EEG and Keystroke fused modality is 93.64% with all features and 88.98% with feature selection. In comparison to EEG, this is 4.9% higher with all features and 10.04% higher with feature selection. In comparison to Keystroke, the improvement is even more (17.42% with all features and 17.86% with feature selection). Thus, it can be summarized that the performance is better with all features and EEG performs better than Keystroke. So, for template matching identification, a large number of features is necessary to achieve good performance.

### E. AUTHENTICATION BY TEMPLATE MATCHING

In the authentication system, the performance is evaluated using EER and area under the ROC curve as shown in Fig. 10. Fig. 10 shows the ROC curve for EEG, Keystroke, and EEG-Keystroke modalities without and with feature selection. The ROC curve is obtained by averaging the ROC curves of all folds for all subjects. The AUC values for ROC curves are 89.23, 87.29, and 93.20 without feature selection and 88.53, 87.74, 93.53 with feature selection, respectively, for EEG, Key, and EEG-Keystroke modalities. Thus, the combination of EEG and Keystroke is outperforming the individual modalities. It is also evident that EEG performs better than Keystroke in template matching. TABLE VI shows the EER for all subjects for EEG, Key, and EEG-Keystroke modalities. It can be seen for TABLE VI that the average EER for EEG, Keystroke, and EEG-Keystroke are 16.66, 16.88, 11.81 without feature selection, and 15.81, 14.48, 10.47 with feature selection, respectively. It can be summarized that the performance improves with feature selection, which is evident from the individual subject's EER value. The lowest EER is achieved for subject 1 in the combination of EEG and Keystroke which is 3.44. The highest EER is seen for subject 2 in Keystroke which is 27.11 without feature selection. Thus, the best-detected subject is Subject 1 and the worst detected subject is Subject 2.

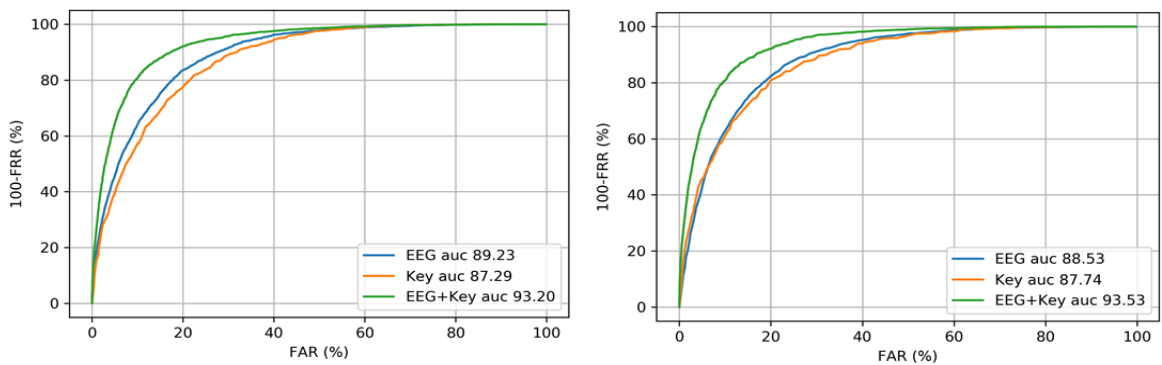

**FIGURE 10.** ROC curves obtained for EEG, Keystroke and EEG-Keystroke modalities: without feature selection (left) and with feature selection (right).

### F. COMPARISON OF COMPUTATIONAL TIME

The average time required for pre-processing and feature extraction of one sample is 184.8 ms. The average training time for classifying 10 subjects is 10.5 s and for 2 subjects, 8.8 s. The mean authentication time of one sample is shown in Table VII for EEG, Keystroke, and EEG-Keystroke modalities. It can be observed from TABLE VII that the lowest times are required in the template matching configuration with feature selection which are 6.33 ms, 4.82 ms, and 7.84 ms for EEG, Keystroke, and EEG-Keystroke, respectively. Template matching is 6 times faster than classification methods on average. Though template matching has slightly lower performances in terms of performance metrics, it outperforms the classification methods in terms of prediction time. So, it is recommended for real-time use.

TABLE VI
EER FOR ALL SUBJECTS IN ALL MODALITIES WITH AND WITHOUT FEATURE SELECTION

| Strategy | Subject | EEG (%) | Keystroke (%) | EEG-Keystroke (%) |
|---|---|---|---|---|
| without feature selection | 1 | 11.93 | 4.60 | **3.75** |
| | 2 | 23.02 | 27.11 | 19.93 |
| | 3 | 19.37 | 19.40 | 15.66 |
| | 4 | 9.73 | 12.71 | 7.42 |
| | 5 | 18.11 | 23.04 | 14.11 |
| | 6 | 19.31 | 29.57 | 16.00 |
| | 7 | 19.35 | 15.00 | 14.00 |
| | 8 | 16.08 | 7.95 | 6.82 |
| | 9 | 12.77 | 17.53 | 9.13 |
| | 10 | 16.95 | 11.84 | 11.24 |
| | Average | **16.66** | **16.88** | **11.81** |
| with feature selection | 1 | 13.75 | 4.40 | **3.44** |
| | 2 | 19.60 | 22.19 | 18.64 |
| | 3 | 15.55 | 16.97 | 10.44 |
| | 4 | 9.28 | 12.95 | 7.17 |
| | 5 | 18.88 | 17.62 | 10.84 |
| | 6 | 18.11 | 26.17 | 16.20 |
| | 7 | 19.22 | 13.06 | 11.91 |
| | 8 | 16.66 | 4.66 | 5.31 |
| | 9 | 12.75 | 15.46 | 10.33 |
| | 10 | 14.80 | 11.35 | 10.48 |
| | Average | **15.81** | **14.48** | **10.47** |

TABLE VII
COMPARISON OF AUTHENTICATION TIME FOR ONE SAMPLE

| Strategy | Method | EEG (ms) | Keystroke (ms) | EEG-Keystroke (ms) |
|---|---|---|---|---|
| without feature selection | Classification | 51.8 | 50.5 | 51.9 |
| | Template Matching | 13.4 | 5.89 | 15.2 |
| with feature selection | Classification | 50.3 | 50.4 | 50.1 |
| | Template Matching | 6.33 | 4.82 | 7.84 |

### G. REAL-TIME EVALUATION

In order to test the multimodal authentication system in a real-world scenario, we designed a GUI-based framework for the real-time processing of EEG signals and keystroke dynamics. The framework utilized the selected machine learning model trained with EEG and keystroke features. Fig. 11 shows the initial GUI for the authentication system.

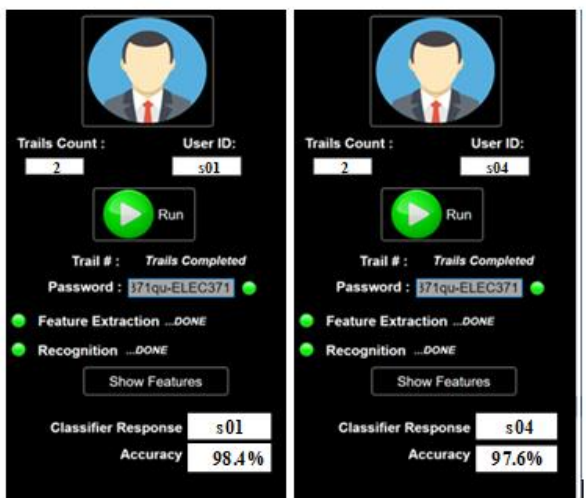

**FIGURE 11. The GUI framework for the real-time evaluation of the proposed multimodal identification and authentication system.**

The first step in this system was to enter the User ID of the test subject and the Trials Count. The trial court was set by the test subject, which indicated the number of times the subject would enter the password. Depending on this value, the data would be segmented and pre-processed accordingly. Furthermore, additional trials resulted in better approximations of the features represented in the input data. The in-house python-based algorithm was used to detect the keypress and key-release intervals for each of the keys inputted by the test subject. The test subject would enter the password for the mentioned number of trials. The feature extraction and recognition will follow and the classifier response would be reported with accuracy. It is observed that for the same number of trials, the authentication and identification of each user can be done through this real-time system with very high accuracy and the accuracy of detection will reach 100% with more than 5 trials for most of the subjects.

## V. DISCUSSION

The key findings, advantages, and disadvantages of this study are discussed below in detail.

The main objective of this study was to build a system that has better anti-spoofing capability (low false-positive rate or high precision) and also higher accuracy and recall (low false-negative rate). From every experimental result of the previous section, it is evident that the proposed multimodal setup has outperformed the individual modalities in terms of accuracy, precision, recall, and F1 score. So, it is clear that our proposed system has overcome the limitation of individual modalities: EEG (low accuracy and high false-negative rate issue) and Keystroke (high false-positive rate).

An important objective of this research was to find out the most suitable fusion scheme for this type of signal or biometric system. From the analysis, it is evident late fusion (summing/multiplying the predicted probabilities of 2 different classifiers for 2 modalities) is the most suitable fusion scheme for this type of system. As EEG and Keystroke produced very different types of features, they are not suitable for any early feature fusion or concatenation. When two separate classifiers are trained on EEG and Keystroke, the classifiers become experts in their corresponding modality, and then in the late fusion stage, they can complement each other. Other fusion methods like – CCA and DCA produced 87.2% and 93.2% accuracy, respectively whereas our method produced 99.9% accuracy.

This research also aimed to build a more interpretable system so that we can understand which features are most important and which parts of the brain are most responsible for producing distinctive features during typing a word. From our analysis, it is evident that the parietal lobe produced the most important features for EEG because it contributes 3 important features ($PZ\_cD2$, $PZ\_cD3$, $PZ\_cD4$) in the top 5 features in the feature ranking of Fig. 6. Most other important features of the top 10 features are related to the frontal lobe (AF3, AF4). From this analysis, it is evident that the area close to the motor cortex is producing important features because, during motor activity (such as typing), the brain area should be active in the motor cortex. Here, it is also evident that wavelet features at different levels are also playing important roles. For example, the wavelet decomposition at levels 2, 3, and 4 are $cD2$, $cD3$, $cD4$, and the top 9 features out of 10 are from these features. This indicates that not all but certain frequency bands of EEG contain biometric traits. As EEG is a non-stationary signal, a composite time-frequency analysis like wavelet analysis plays a crucial role. The reason is, wavelet analysis identifies not only the slowly varying components but also the abruptly changing components of a non-stationary signal. In any non-stationary signal, the frequency can change with respect to time. Wavelets can identify the frequency components in each time bin which is not possible for time or frequency domain alone. For this reason, a comprehensive analysis can be done and distinctive patterns of non-stationary signals can be identified by wavelets.

The experimental results show that ensemble methods with bagging (RnF) and boosting (XGBoost) have performed well in most cases. In both cases, *N* learners are created from 1 learner. This method matches the ensembling of multiple modalities. That is why these classifiers are chosen and they show better performance. Other reasons are that both classifiers reduce variance and thus overfitting. In reducing overfitting, it was evident that bagging outperforms boosting. Both classifiers make the final decision by averaging the *N* learners but bagging applies an equally-weighted average but boosting applies a non-equally weighted average. Equally weighted average matches the score level fusion method and thus bagging performs better than boosting.

Although our proposed multimodal authentication system achieved very high accuracy and reliability with additional layers of security, there are still a few areas that can be improved. The models were trained on EEG and keystroke data from 10 different users. The performance of the model is yet to be tested on a rich dataset containing inputs from a large number of users. Another limitation of this study is the use of only one specific password for all users. This authentication system can be extended by implementing several distinct passwords into the multimodal dataset and learning algorithms. Even though EEG biometrics and keystroke dynamics have been integrated into an existing password/PIN framework, the requirement of the EEG headset may not appeal to finance or retail sectors. But with the advancement in technologies, it may be possible soon to integrate EEG biometrics into wireless portable earphones. For such cases, the proposed authentication system can be evaluated using a single channel or dual channel EEG data, along with keystroke dynamics from smartphones. This will enable faster adoption of the authentication system into mainstream devices. Finally, in this study, the features from the data were manually handpicked, and then machine learning techniques were applied to find a pattern in the dataset. This process can be automated using an auto-encoder or convolutional neural network (CNN), which might enhance the real-time system performance from single-trial evaluation.

## VI. CONCLUSION

In this paper, we demonstrated a multi-modal system that merged EEG and keystroke dynamics for user identification and authentication. We integrated our multimodal system into the widely used password/PIN-based authentication system, which reduces the inconvenience of the end-users as they do not need to adapt to a new authentication process. Using the features extracted from the input data, the behavioral pattern can be used to identify the user with a confidence level, and the password can be used for authorization. We implemented a variety of machine learning techniques on the pre-processed data and selected a model with the highest accuracy and other performance metrics. Furthermore, we trained separate tailored models for each user to further improve the identification performance of the multimodal system. The personalized models were used for classifying users as genuine or impostors. We also developed a fast binary template matching method for identification and authentication. To the best of our knowledge, this is the first implementation of EEG biometrics and Keystroke dynamics merged into a multimodal system for identification and authorization. This implementation provides additional layers of security through the EEG signal, which is difficult to mimic, and Keystroke dynamics, which captures the behavior and rhythm of the user's typing with high accuracy. Despite the benefits, we have discussed a few limitations of our system, which we will address in our future studies. Furthermore, we believe that our demonstration will encourage further studies in such multimodal systems, which can then be deployed to different sectors for real-world application.

### FUNDING
This work was made possible by Qatar National Research Fund (QNRF) NPRP12S-0227-190164 and UREP23-027-2-012. The statements made herein are solely the responsibility of the authors.

### FUNDING
This work was made possible by Qatar National Research Fund (QNRF) NPRP12S-0227-190164 and UREP23-027-2-012. The statements made herein are solely the responsibility of the authors.

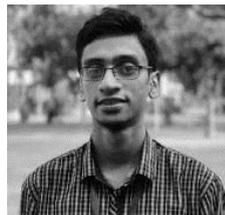

**ARAFAT RAHMAN** received his B.Sc. (Eng.) degree from the Department of Electrical and Electronic Engineering at the University of Dhaka, Bangladesh, and currently doing M.Sc. in the Department of Biomedical Physics and Technology (BPT) at the University of Dhaka, Bangladesh. He is also currently working as a research student at the University of Dhaka, Bangladesh. His current research interest includes biomedical image and signal processing, machine learning, activity recognition, and biometrics. He has expertise in designing and developing machine learning and deep learning pipeline for various signals like EEG, EMG, and ECG. He has also expertise in designing EMG controlled prosthetic hand, 3D printing, Laser Cutting, developing electrocardiogram (ECG), electromyogram (EMG) circuit, developing Howland constant current source and Instrumentation amplifier to measure tetra polar bio-impedance, Detection of the different stage of brain activity by analyzing various EEG wave, and accelerometer-based human activity recognition. He has published several conference papers that are related to EMG-controlled prosthetic hand and accelerometer-based human activity recognition. He achieved an excellent paper award at the 4th Int. Conf. on Imaging, Vision and Pattern recognition, Japan, Aug 2020 and placed 3[rd] position in the 2nd Nurse Care Activity Recognition Challenge, at HASCA Workshop, 2020 ACM International Symposium on Wearable Computers (UbiComp/ISWC'20), Mexico, Sept. 2020.

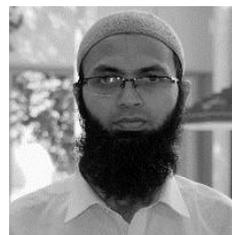

**MUHAMMAD E. H. CHOWDHURY** (Senior Member, IEEE) received the B.Sc. and M.Sc. degrees with record marks from the Department of Electrical and Electronics Engineering, University of Dhaka, Bangladesh, and the Ph.D. degree from the University of Nottingham, U.K., in 2014. He worked as a Postdoctoral Research Fellow and a Hermes Fellow at the Sir Peter Mansfield Imaging Centre, University of Nottingham. He is currently working as an Assistant Professor with the Department of Electrical Engineering, Qatar University. Before joining Qatar University, he worked in several universities in Bangladesh. He has two patents and published around 80 peer-reviewed journal articles, conference papers, and four book chapters. His current research interests include biomedical instrumentation, signal processing, wearable sensors, medical image analysis, machine learning, embedded system design, and simultaneous EEG/fMRI. He is currently running several NPRP and UREP grants from QNRF and internal grants from Qatar University along with academic and government projects. He has been involved in EPSRC, ISIF, and EPSRC-ACC grants along with different national and international projects. He has worked as a Consultant for the projects entitled Driver Distraction Management Using Sensor Data Cloud (2013–14, Information Society Innovation Fund (ISIF) Asia). He is an Active Member of British Radiology, the Institute of Physics, ISMRM, and HBM. He received the ISIF Asia Community Choice Award 2013 for a project entitled Design and Development of Precision Agriculture Information System for Bangladesh. He has recently won the COVID-19 Dataset Award and National AI Competition awards for his contribution to the fight against COVID-19. He is serving as an Associate Editor for IEEE Access and a Topic Editor and Review Editor for Frontiers in Neuroscience.

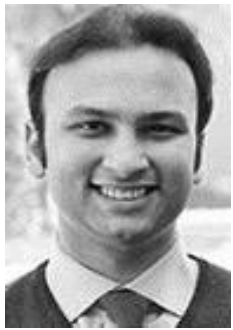

**AMITH KHANDAKAR** (Senior Member, IEEE) received a B.Sc. degree in electronics and telecommunication engineering from North South University, Bangladesh, and a master's degree in computing (networking concentration) from Qatar University, in 2014. He graduated as the Valedictorian (President Gold Medal Recipient) of North South University. He is currently the General Secretary of the IEEE Qatar Section and also the Qatar University IEEE Student Branch Coordinator and an Adviser (Faculty). He is also a certified Project Management Professional and the Cisco Certified Network Administrator. He was a Teaching Assistant and Lab Instructor for two years for courses, such as mobile and wireless communication systems, the principle of digital communications, introduction to communication, calculus and analytical geometry, and Verilog HDL: modeling, simulation, and synthesis. Simultaneously, he was a Lab Instructor for the following courses: programming course "C," Verilog HDL, and general physics course. He has been with Qatar University, since 2010. After graduation, he was a Consultant in a reputed insurance company in Qatar and in a private company that is a sub-contractor to National Telecom Service Provider in Qatar.

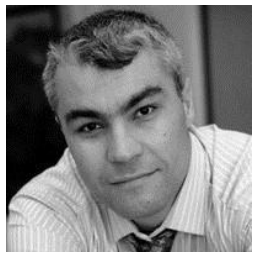

**SERKAN KIRANYAZ** joined Qatar University in 2015 as a full professor in the Department of Electrical Engineering. During 2010-2015, he was a Professor in the Department of Signal Processing of Tampere University of Technology. He published 2 books, 3 book chapters, more than 100 journal papers in many IEEE Transactions and other high-impact journals, and around 90 papers in international conferences. His principal research field is "Intelligent Systems" which comprises systems, devices, algorithms, software, and processes requiring intelligent processing of, possibly "Big" data and sensor signals together with artificial reasoning. He made significant contributions to bio-signal analysis, particularly ECG analysis and processing, classification and segmentation, signal processing and management, computer vision, scene analysis, automatic object segmentation, and machine learning strategies with applications to recognition, classification, detection, and evolutionary machine learning, swarm intelligence, stochastic optimization, and deep learning. Particularly, they are among the pioneers in patient-specific ECG classification. His work became one of the the-most-read articles in IEEE Transaction on Biomedical Engineering and has now more than 150 citations according to Google Scholar. His most recent work on real-time patient-specific ECG classification too became one of the most popular articles in the IEEE Transaction on Biomedical

Engineering. Prof. Kiranyaz and Prof. Gabbouj with their research team won 2nd place in the PhysioNet Challenge Grand Competition, where 48 international teams worldwide competed to achieve the most accurate system for the classification of Normal/Abnormal heart sound recordings. Such extensive collaboration and prior research works will serve as the basis required for this project and this collaboration shall continue with also researcher exchanges between QU, TUT, and IUE.

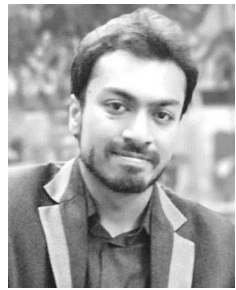

**KH SHAHRIYA ZAMAN** received his B.Sc. degree in Electrical and Electronic Engineering from Stamford University Bangladesh in 2013. He received his M.Sc. degree in Electrical and Electronic Engineering from Coventry University, UK, in 2016. He is currently pursuing a Ph.D. degree with the Department of Electrical, Electronic and Systems Engineering, Universiti Kebangsaan Malaysia, where he is involved in developing specialized hardware architectures for artificial neural networks. His current research interests include deep learning, hardware accelerators for deep learning, and custom hardware architectures for low power devices.

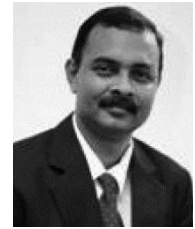

**MAMUN BIN IBNE REAZ** (SM'02) was born in Bangladesh in December 1963. He received the B.Sc. and M.Sc. degrees in applied physics and electronics from the University of Rajshahi, Bangladesh, in 1985 and 1986, respectively, and the D.Eng. degree from Ibaraki University, Japan, in 2007. He is currently a Professor with the Department of Electrical, Electronic and Systems Engineering, Universiti Kebangsaan Malaysia, Malaysia, involving in teaching, research, and industrial consultation. He has been a Regular Associate of the Abdus Salam International Centre for Theoretical Physics, since 2008. He has vast research experience in Japan, Italy, and Malaysia. He has published extensively in the area of IC Design and Biomedical application IC. He has authored or coauthored more than 200 research articles in design automation and IC design for biomedical applications. He was a recipient of more than 50 research grants (national and international).

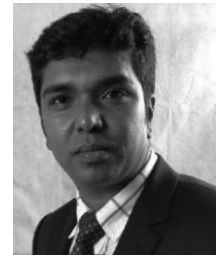

**MOHAMMAD TARIQUL ISLAM** (Senior Member, IEEE) is currently a Professor with the Department of Electrical, Electronic and Systems Engineering, Universiti Kebangsaan Malaysia (UKM), and a Visiting Professor with the Kyushu Institute of Technology, Japan. He is the author or co-author of about 500 research journal articles, nearly 175 conference papers, and a few book chapters on various topics related to antennas, microwaves, and electromagnetic radiation analysis with 20 inventory patents filed. Thus far, his publications have been cited 5641 times and his H index is 38 (Source: Scopus). His Google scholar citation is 8200 and his H-index is 42. His research interests include communication antenna design, satellite antennas, and electromagnetic radiation analysis. He is a Chartered Professional Engineer (C.Eng.), a member of IET, U.K., and a Senior Member of IEICE, Japan. He was a recipient of the 2018 IEEE AP/MTT/EMC Excellent Award, the Publication Award from the Malaysian Space Agency, in 2014, 2013, 2010, and 2009, and the Best Paper Presentation Award from the International Symposium on Antennas and Propagation (ISAP 2012), Nagoya, Japan, in 2012, and IconSpace, in 2015. He received several international gold medal awards, including the Best Invention in Telecommunication Award, the Special Award from Vietnam for his research and innovation, and best researcher awards at UKM, in 2010 and 2011. He also won the Best Innovation Award, in 2011, and the Best Research Group in ICT Niche by UKM, in 2014. He was a recipient of more than 40 research grants from the Malaysian Ministry of Science, Technology and Innovation, the Ministry of Education, the UKM Research Grant, and international research grants from Japan and Saudi Arabia. He serves as a Guest Editor for Sensors Journal and an Associate Editor for IEEE Access and Electronics Letters (IET).

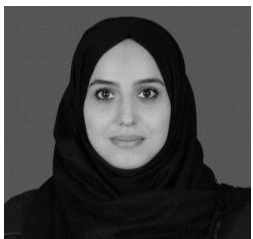

**MAYMOUNA EZEDDIN** received her B.Sc in electrical engineering and minor in business for a non-business student from Qatar University, Qatar in 2020. She is currently studying M.Sc Data Science and Engineering at Hamad Bin Khalifa University, Qatar, and works as Research Assistant at the College of Engineering, Qatar University. She worked on several Undergraduate research projects (UREP) funded by Qatar national research fund (QRNF). Her research interests include signal processing and machine learning.

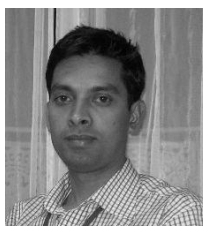

**MUHAMMAD ABDUL KADIR** received the B.Sc. (Honours) and Master of Science degrees in Physics from the University of Dhaka in Bangladesh. He was a visiting postgraduate researcher at the University of Warwick, United Kingdom as a Commonwealth Scholar. MAK completed his Ph.D. in Biomedical Physics jointly from the University of Dhaka and the University of Warwick. He is a life member of Bangladesh Physical Society and Bangladesh Medical Physics Association. He is working as an associate professor at the Department of Biomedical Physics & Technology of the University of Dhaka, Bangladesh since 2018. His research focuses on biomedical instrumentation and medical applications of electrical impedance techniques. He has practical experience in the design and development of multi-frequency electrical bioimpedance measurement systems. He is also interested in biomedical signal and image analysis and machine learning techniques for disease diagnosis.

## Supplementary Materials

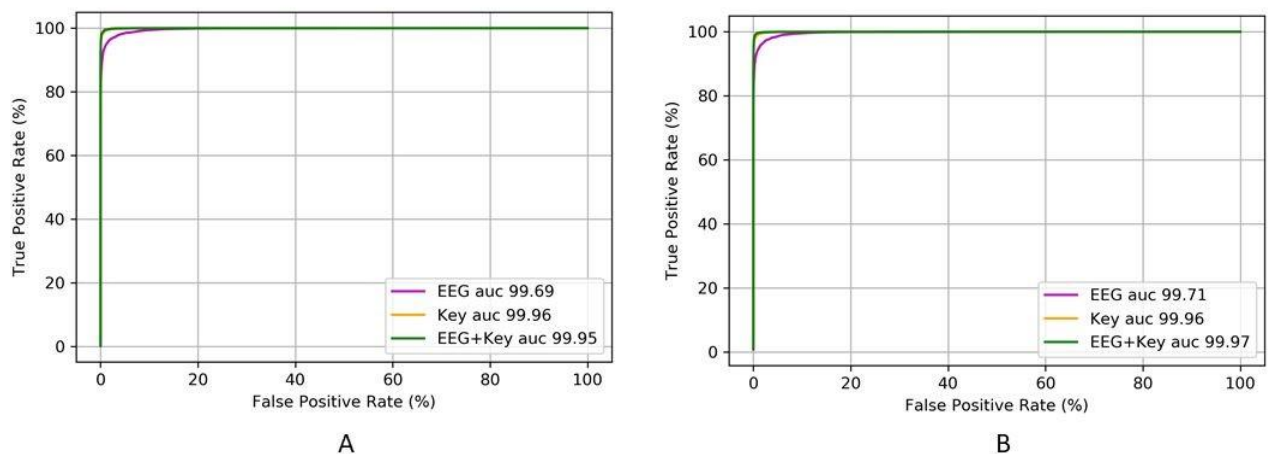

**Supplementary Figure 1**: ROC curves for the best classifiers using independent modality and fused modalities: without feature selection (A) and with feature selection (B).

**Supplementary Table 1:** Performance evaluation for personalized classifications using augmented raw EEG data

| Augmentation | Subject | Accuracy | Precision | Recall | F1 score |
|---|---|---|---|---|---|
| TimeW | 1 | 98.91 | 98.43 | 95.39 | 96.85 |
|  | 2 | 96.72 | 98.23 | 83.50 | 89.22 |
|  | 3 | 99.14 | 99.51 | 95.51 | 97.41 |
|  | 4 | 99.46 | 99.67 | 97.30 | 98.29 |
|  | 5 | 97.71 | 98.17 | 88.94 | 92.94 |
|  | 6 | 97.53 | 98.65 | 87.50 | 92.17 |
|  | 7 | 98.71 | 99.29 | 93.51 | 96.17 |
|  | 8 | 98.57 | 98.66 | 92.94 | 95.58 |
|  | 9 | 99.11 | 99.02 | 95.94 | 97.42 |
|  | 10 | 98.34 | 99.07 | 91.52 | 94.89 |
|  | **Average** | **98.42** | **98.85** | **92.25** | **95.12** |
| Jitter | 1 | 99.10 | 98.56 | 96.39 | 97.44 |
|  | 2 | 96.90 | 97.68 | 84.94 | 90.09 |
|  | 3 | 98.80 | 97.91 | 95.33 | 96.57 |
|  | 4 | 99.80 | 99.89 | 99.00 | 99.44 |
|  | 5 | 98.11 | 98.42 | 90.94 | 94.29 |
|  | 6 | 97.52 | 98.05 | 87.94 | 92.25 |
|  | 7 | 98.81 | 98.84 | 94.44 | 96.51 |
|  | 8 | 98.92 | 98.90 | 94.94 | 96.82 |
|  | 9 | 99.11 | 98.56 | 96.39 | 97.44 |
|  | 10 | 98.73 | 99.29 | 93.51 | 96.17 |
|  | **Average** | **98.58** | **98.61** | **93.38** | **95.70** |